\documentclass{aastex} 

\usepackage{emulateapj5}
\usepackage{latexsym}

\received{26 March 2001}
\accepted{16 April 2001}
\slugcomment{The Astrophysical Journal Letters}
 
\lefthead{Guerrero et al.}
\righthead{X-ray Points Sources in NGC\,6543 and NGC\,7293}

\begin{document}

\title{The Enigmatic X-ray Point Sources at the Central Stars of \\
       NGC\,6543 and NGC\,7293}

\author{Mart\'{\i}n A.\ Guerrero\altaffilmark{1}, 
        You-Hua Chu\altaffilmark{1}, 
        Robert A. Gruendl\altaffilmark{1}, \\ 
        Rosa M. Williams\altaffilmark{2, 3}, 
        and James B. Kaler\altaffilmark{1}}
\altaffiltext{1}{Astronomy Department, University of Illinois, 
        1002 W. Green Street, Urbana, IL 61801, USA;
        chu@astro.uiuc.edu, mar@astro.uiuc.edu, gruendl@astro.uiuc.edu,
        kaler@astro.uiuc.edu}
\altaffiltext{2}{National Research Council Associate} 
\altaffiltext{3}{NASA's GSFC, code 662, Greenbelt, MD 20771, USA;
        rosanina@lhea1.gsfc.nasa.gov}

\submitted{{\it Received 2001 March 26; accepted 2001 April 16}}

\begin{abstract}

Recent {\it Chandra} ACIS-S observations have detected a point source 
at the central star of NGC\,6543 and confirmed the point source nature 
of the hard X-ray emission from NGC\,7293.  
The X-ray spectra of both sources peak between 0.5 keV and 1.0 keV
and show line features indicating a thin plasma at temperatures of a 
few times 10$^6$ K.
Their X-ray luminosities are $10^{30}$ erg~s$^{-1}$ and 
$3\times10^{29}$ erg~s$^{-1}$, respectively.  
We have considered four different mechanisms to explain the nature
of these sources.
The X-ray emission from the central star of NGC\,6543 may originate
from the coronal activity of an undetected companion star or from
shocks in its fast stellar wind, while the hard X-ray emission from
NGC\,7293 might be ascribed to an undetected dMe companion.
Follow-up observations are needed to determine the existence and 
natures of these stellar companions.  

\end{abstract}

\keywords{planetary nebulae: general -- planetary nebulae: individual
(NGC\,6543, NGC\,7293) -- stars: AGB and post-AGB -- X-rays: stars}

\section{Introduction}

Central stars of planetary nebulae (PNs) are expected to emit soft X-rays 
if their effective temperatures are greater than 100,000 K.  
Indeed, such X-ray sources have been detected by {\it Einstein}, {\it 
EXOSAT}, and {\it ROSAT}, and the X-ray photons detected have energies 
well below 0.5 keV \citep{Hoare95,GCG00}.  
In addition to the soft X-ray emission from the photosphere, {\it ROSAT} 
observations of the central stars of LoTr\,5 and NGC\,7293 (also known 
as the Helix Nebula) show a harder X-ray component peaking between 0.5 
and 1.0 keV \citep{leahy94,GCG00}. 
The central star of LoTr\,5 is a known binary \citep{FK83}, and the 
coronal activity of its G5~III companion may be responsible for the 
hard X-ray emission \citep{Hoare95,Jas96}.  
The central star of NGC\,7293, on the other hand, is not a known binary
\citep{C99}.  

It has been suggested that the hard X-ray emission from NGC\,7293 originates 
from the interaction between the fast stellar wind and the ambient nebular 
material \citep{leahy94,leahy96}.  
If this is true, the X-ray emission from NGC\,7293 should be spatially 
extended, similar to that observed in NGC\,6543 \citep{CGGWK01}.  
We have obtained {\it Chandra} observations of NGC\,7293 and find that 
this source remains unresolved by {\it Chandra}'s high resolution.  

We have also obtained {\it Chandra} observations of NGC\,6543 (the 
Cat's Eye Nebula) and find that its central star emits X-rays, too 
\citep{CGGWK01}.  
The X-ray spectrum of this source peaks at $>$ 0.5 keV; 
thus, NGC\,6543 joins NGC\,7293 as hosts of enigmatic point X-ray sources.  
This paper reports our {\it Chandra} observations of the point 
sources in NGC\,6543 and NGC\,7293.  
The observations are described in \S2, the properties of the X-ray 
emission are presented in \S3, {the~possible} origins of these X-ray
sources are discussed in \S4, and the conclusions are summarized in \S5. 

\section{Observations}

NGC\,6543 and NGC\,7293 were observed with the Advanced CCD Imaging 
Spectrometer (ACIS) on board {\it Chandra X-ray Observatory} on 2000 
May 10--11 and 1999 November 17--18 for total exposure times of 46.0 
and 47.7 ks, respectively.  
The nebular centers were placed on the back-illuminated (BI) CCD chip 
S3, the nominal aim point for the ACIS-S array.
The instrumental FWHM of the ACIS observations is 0\farcs7--0\farcs8 
at $\le$ 1 keV \citep{WMcN99}.
The energy resolution, $E/\Delta E$, is $\sim 4.3$ at 0.5 keV and $\sim 
9$ at 1.0 keV.

We received Level 1 and Level 2 processed data from the {\it Chandra}
Data Center.  The data reduction and analysis were performed using the
{\it Chandra} X-ray Center software CIAO V1.1.5 and HEASARC FTOOLS and
XSPEC V11.0.1 routines \citep{Arnaud96}.
No background ``flares" affected the observations and no time intervals 
needed to be removed.

\section{Properties of the X-ray Emission}

\subsection{Identification of Point Sources}

The $Chandra$ observations of NGC\,6543 detected 1,950$\pm$50 counts, 
of which $\sim$100 counts originated at a point-like source and the rest 
corresponded to diffuse emission \citep{CGGWK01}.  
The observations of NGC\,7293 detected 2,100$\pm$50 counts, all from a 
point-like source.  
The {\it Chandra} coordinates of both point-like sources  
are within 1$''$ of the coordinates of the central stars 
derived from the Digitized Sky Survey\footnote{
The Digitized Sky Survey was produced at the Space Telescope Science 
Institute under US government grant NAGW-2166 based on photographic 
data obtained using the UK Schmidt Telescope and the Oschin Schmidt 
Telescope on Palomar Mountain.}. 
To further examine their relative positions, we have matched field 
stellar sources detected in the ACIS 
observations with bright stars in the Digitized Sky Survey.  
When the field sources are aligned, we find that the agreement in position
between the point-like X-ray source and the central star is better than
0\farcs5 for both NGC\,6543 and NGC\,7293.  
Therefore, we conclude that these point-like X-ray sources are coincident 
with the central stars.

It is difficult to assess whether the point-like X-ray source at 
NGC\,6543's central star is truly unresolved because only $\sim$100 
counts are detected and it is superposed on bright diffuse 
{X-ray~emission.}  
This point-like X-ray source cannot be a local peak of the diffuse 
emission {because~they~have~different} spectral shapes (see \S3.2); 
thus, it is likely a true point source associated with 
the {central~star.}

The ACIS observations of NGC\,7293 have detected adequate counts to 
allow an accurate comparison between the surface brightness profile 
of the point-like source and the instrumental point spread function.  
The FWHM of the surface brightness profile of the X-ray source in 
NGC\,7293 is 0\farcs8, comparable to the instrumental FWHM. 
Therefore, the X-ray source in NGC\,7293 is unresolved, and is a true
``point" source.

\subsection{Spectral Properties of NGC\,6543}

The spectrum of the point source at the central star of NGC\,6543 was 
extracted using a round source aperture of 1\farcs1 radius and a
concentric annular background aperture of radii from 1\farcs5 to 
2\farcs5.  
The background region was selected to be representative of the diffuse 
emission in the central region of the nebula while avoiding the bright 
emission along the rim \citep[see image of diffuse emission in][]{CGGWK01}.  
The background-subtracted spectrum of the central star of NGC\,6543 is 
presented in Fig.~1.  
It shows emission-line features at 0.45--0.7 keV and $\sim0.9$ keV, 
and much fainter emission at higher energies in the range
of 1.0 to 2.0 keV. 
The feature between 0.45 and 0.7 keV is brighter and wider, and 
corresponds to the He-like triplet of \ion{O}{7} at $\sim$0.57~keV 
and the H-like Ly$\alpha$ line of \ion{O}{8} at 0.65~keV.
The feature at $\sim0.9$ keV is fainter and may be attributed to
the He-like triplet of \ion{Ne}{9} at $\sim$0.92~keV.  
The spectrum at 1.0--2.0 keV has too few counts for us to determine 
whether it is line or continuum emission.

The line features in the spectrum of the central star of NGC\,6543 are 
consistent with thermal plasma emission. 
The number of counts is too small for reliable spectral fits; 
therefore, we will limit our analysis to the comparison of the observed 
spectral shape with thermal plasma emission models convolved with an 
absorption column and the instrumental response.  
For this, we used the MEKAL model in the XSPEC software package 
\citep{MLvdO86,K92,L92}, and an absorption column density, $N_{\rm H} = 
8\times$10$^{20}$ cm$^{-2}$, determined from the ACIS-S spectral fits 
of the diffuse emission from NGC\,6543 \citep{CGGWK01}.
Solar abundances are adopted for both the emitting plasma and 
absorbing material.

Several models have been considered to simulate the observed spectrum
(see Fig.~1).
For models with one temperature component, the relative intensity of 
the emission features at 0.45--0.7 keV and $\sim0.9$ keV can be 
simulated only if the abundance of Ne is greatly enhanced from the 
solar value; 
the best model has a plasma temperature of $T \sim$ 2$\times$10$^6$~K.
For models with two temperature components, no anomalous Ne abundances 
are needed, and the observed spectral shape can be described by a MEKAL 
model with $T_1 \sim 2\times10^6$~K and $T_2 \sim 9\times10^6$~K.  
The unabsorbed X-ray luminosity, calculated from either 
the best one- or two-temperature model, is $\sim10^{30}$ erg~s$^{-1}$
in the 0.3--2.0 keV band, for a distance of 1 kpc \citep{Reed99}.  

\subsection{Spectral Properties of NGC\,7293}

The spectrum of the point source at the central star of NGC\,7293 was 
extracted using a round source aperture of 1\farcs5 radius and a
concentric annular background aperture of radii from 10\arcsec\ to 
25\arcsec. 
The background-subtracted spectrum (Fig.~2a) shows emission from 0.3 
to 2.0 keV with numerous line features (note that ACIS is not 
sensitive to the soft X-ray component detected by $ROSAT$).  
The emission in the energy range of 0.5--0.7 keV consists of the 
\ion{O}{7} lines at 0.57 keV and \ion{O}{8} line at 0.65 keV.
Note that the relative intensities of the \ion{O}{7} and \ion{O}{8}
lines of NGC\,7293 are reversed from those of NGC\,6543, indicating 
a higher plasma temperature for the source in NGC\,7293.
The peak emission, in the energy range of 0.7--0.85 keV, corresponds
to the \ion{Fe}{17} lines at $\sim$0.73 keV and $\sim$0.83 keV. 
The secondary peak, at 0.9 keV, is attributed to \ion{Ne}{9} 
lines at $\sim$0.92 keV.  
The \ion{Ne}{10} line and Fe\,L blend at 1.02 keV and 1.10 keV, 
respectively, can also be identified in the spectrum.  
In the energy range above 1.1 keV, the emission drops off steadily;
the low S/N ratio prevents us from identifying the emission lines
unambiguously.

The spectrum of the point source in NGC\,7293 has been fit using the 
MEKAL model.  
The best fit (see Fig.~2a) has $T \simeq 7.4\times$10$^6$~K and 
$N_{\rm H} \simeq 4\times10^{20}$ cm$^{-2}$.  
The goodness of the fit can be assessed from the reduced $\chi^2$ 
plotted as a function of $N_{\rm H}$ and $kT$ (Fig.~2b).
The 99\% confidence contour spans $N_{\rm H}$ = 
2--8$\times$10$^{20}$~cm$^{-2}$ and $T$ = 7--8$\times$10$^6$ K
(or $kT$ = 0.60--0.67 keV).  
This absorption column density is, within the error limit, consistent 
with that determined from $ROSAT$ PSPC observations, 
0--2.8$\times$10$^{20}$ cm$^{-2}$ \citep{leahy96}.
The unabsorbed X-ray luminosity is $\sim$3$\times$10$^{29}$ erg~s$^{-1}$ 
in the 0.3--2.0 keV band, for a distance of 210 pc \citep{HDMP97}.  

\subsection{Temporal Brightness Variation of the Point Source in NGC\,7293}

The 47.7 ks observations of NGC\,7293 were made in two intervals,
36.7 ks and 11.0 ks, separated by a 22.6 ks gap.
We have examined the temporal variation of the brightness of the
source over this time span by dividing the observations into nine 
roughly equal time bins.
The count rate is plotted against time in Fig.~3.
The average ACIS count rate of this source is consistent with 
that expected from the {\it ROSAT} PSPC observations of the 
hard component on 1992 May 12--13. 
The ACIS count rates from the seven bins in the first time interval
show 1$\sigma$ level variations.
The count rates from the final two bins, on the other hand, show a 
noticeable decrease from the first time interval.
Both the peak-to-peak variation and the difference between the
mean of the first seven bins and the mean of the last two bins 
have greater than 3$\sigma$ significance.
These results provide the first detection of short-term 
variability of the X-ray source in NGC\,7293.

\section{Possible Origins of the X-ray Emission}

\subsection{Wind-Nebula Interaction}

It has been suggested that the hard X-ray emission from the central
star of NGC\,7293 originates from the interaction between its 
stellar wind and the surrounding nebula \citep{leahy96}, because 
central stars of PNs often exhibit energetic fast stellar winds 
\citep{PP91}. 
Such interacting-wind models \citep{KPF78} show that the PN interior 
is filled with hot gas (the shocked fast wind), and the X-ray emission 
from the hot gas will peak near the inner wall of the PN shell, where 
the density of the hot gas is raised by the mass evaporation across 
the interface \citep[e.g.,][]{MF95,ZP98}.
Diffuse X-ray emission compatible with this picture has been detected 
by {\it Chandra} in the interiors of BD+30$^\circ$3639, NGC\,6543, 
and NGC\,7027 (Kastner et al.\ 2000, 2001; Chu et al.\ 2001). 
In all cases, the observed spectra are consistent with a thin plasma 
at a few $\times$10$^6$ K.

The ACIS spectra of the central stars of NGC\,6543 and NGC\,7293 
are qualitatively similar to those expected in the wind-nebula 
interaction models, but the X-ray emission is confined to the 
vicinity of the central star as opposed to peaking near the inner 
wall of the PN shell.  
Furthermore, the central star of NGC\,7293 currently does not seem 
to possess a measurable fast stellar wind \citep{CSP85}.
It is unlikely that the X-ray emission from these point sources 
originates from wind-nebula interaction.  

\subsection{Shocks in the Stellar Wind}

Shocks in fast stellar winds are believed to be responsible for
the X-ray emission from massive O and B stars (Cassinelli et al.\ 
1994). 
X-ray observations of O and B stars show that their spectra are
consistent with thin plasma emission at a few times 10$^6$ K and 
that $L_{\rm X}/L_{\rm bol} \approx 10^{-7}$ \citep{Cetal89}.
While the central star of NGC\,7293 does not have a measurable
fast wind, the central star of NGC\,6543 does exhibit a strong 
fast stellar wind \citep{PP91}.
With an $L_{\rm bol}$ of $\sim10^{37}$ ergs~s$^{-1}$ \citep{PCL89},
the $L_{\rm X}/L_{\rm bol}$ of the central star of NGC\,6543 is 
within the range expected for sources with wind shocks.
Therefore, the X-ray emission from the central star of NGC\,6543
may originate from shocks in its fast stellar wind.

\subsection{Accretion of Material from a Close Binary Companion}

A compact object with a close binary companion may form an accretion 
disk and emit X-rays (e.g., the low- and high-mass X-ray binaries).  
The maximum temperature that can be achieved in the accretion disk 
material scales with $M^{1/4} \dot M^{1/4} R^{-3/4}$, where $M$ and 
$R$ are the mass and radius of the accreting star, and $\dot M$ is 
the mass transfer rate in the binary system \citep{P81}.
For a radius of 0.01 R$_\odot$ and a mass of 0.6 M$_\odot$, typical
for PN central stars, the maximum temperature achievable in the 
accretion disk is well below 10$^5$ K for a mass transfer rate of 
$< 10^{-7}$ M$_\odot$~yr$^{-1}$.
No X-ray emission is expected from such cold accretion disks.

Alternatively, the accreted mass may fall onto the surface of a 
compact object directly and emit X-rays.
For white dwarfs, the infall velocity is so high that a shock 
develops above the star and the infalling gas is shock-heated 
to 10$^8$ K and emits X-rays, e.g., dwarf novae, polars, and
intermediate polars \citep{S99,C90,P94}.
The X-ray spectra of the central stars of NGC\,6543 and NGC\,7293 
are clearly too soft;
furthermore, the optical brightness does not show the large variability 
observed in white dwarfs that accrete mass at high rates (A.\ Landolt, 
private communication).  
Accretion of material from a close binary companion cannot contribute 
to the observed X-ray emission from these point sources.  

\subsection{Coronal Activity of the Central Star or a Dwarf Companion}

X-ray emission from stellar coronae has similar spectral properties to 
the X-ray emission from the central stars of NGC\,6543 and NGC\,7293.  
Coronal activity is powered by convection and differential rotation in 
the envelopes of late type F--K stars and dMe flare stars (dwarf M stars 
with emission lines).  
Evolved stars, going from the end of the AGB phase to the proto-PN stage, 
may also have convective envelopes \citep{B01}.  
As a star evolves into a white dwarf, however, the ionization of H and He 
in its photosphere increases, and the convection and coronal activity will 
cease at $T_{\rm eff} > 30,000$ K \citep{BC71}. 
Neither NGC\,6543 nor NGC\,7293 are proto-PNs.  
In fact, the central star of NGC\,7293 is already a 100,000~K hot white 
dwarf.  
Therefore, it is unlikely that they have coronal activity to produce
the observed X-ray emission.

It is possible that the central stars of NGC\,6543 and NGC\,7293 have 
binary companions with coronal activity.  
The luminous central star of NGC\,6543 \citep[$\sim 5,600$ 
L$_\odot$,][]{PCL89} can easily hide a F--M dwarf companion.  
An undetected companion has been suggested to explain the precessing 
collimated outflows in NGC\,6543 \citep{MS92,HB94}.  
A careful search is needed to determine whether a binary companion 
exists.  

Observations of the central star of NGC\,7293 have shown that no 
companion star with spectral type earlier than M5 is present 
\citep{C99}.  
On the other hand, the temporal variability of NGC\,7293 is similar 
to those observed from dMe flare stars in quiescent state \citep{W96}.  
Furthermore, the broad ($\sim 300$ km~s$^{-1}$) variable H$\alpha$ 
emission from the central star of NGC\,7293 recently detected by 
\citet{G01} is similar to that seen during flares of dMe stars.  
A dMe companion is currently the most plausible explanation for 
the X-ray emission from NGC\,7293.

\section{Conclusions}

Using {\it Chandra} ACIS-S observations, we have discovered a point 
source at the central star of NGC\,6543 and confirmed the point source 
nature of the hard X-ray emission from NGC\,7293.  
The spectra show line features indicating a thin plasma at 
temperatures a few times 10$^6$ K.  
The luminosities of these sources are $10^{30}$ erg~s$^{-1}$ and 
$3\times10^{29}$ erg~s$^{-1}$, respectively.  
The central point source in NGC\,6543 is detected because of  
the unprecedented resolution and sensitivity of $Chandra$, 
and the central star of NGC\,7293 is detected because it 
one of the nearest PNs.  
We conclude that the X-ray emission from the central star of NGC\,6543 
may be ascribed to coronal activity of a late-type companion star or to 
shocks in its stellar wind;   
the X-ray emission from the central star of NGC\,7293 may originate 
from a companion dMe star.  
Follow-up observations are needed to determine the existence and natures 
of these stellar companions.

\acknowledgements
This work is supported by the {\it Chandra} X-Ray Observatory Center 
Grant Number GO0-1004X.  MAG is supported partially by the Direcci\'on 
General de Ense\~nanza Superior e Investigaci\'on Cient\'{\i}fica of 
Spanish Ministerio de Educaci\'on y Cultura.

\newpage

\begin{figure}
\epsscale{0.9}
\centerline{\plotone{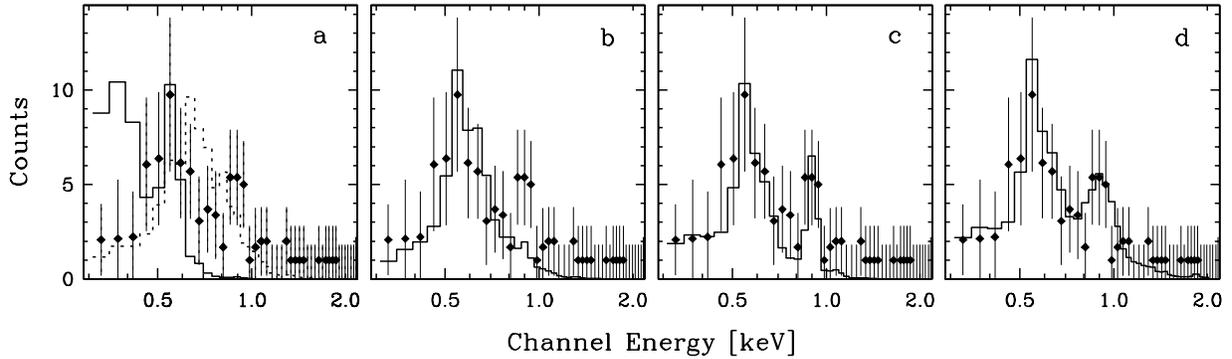}}
\caption{
{\it Chandra} ACIS-S spectrum of the central star of NGC\,6543
overplotted by different MEKAL thin plasma emission models with:
({\it a}) $T$ = $1\times10^6$~K ({\it solid histogram}) 
    and $T$ = $3\times10^6$~K ({\it dotted histogram}), and solar 
    abundances; 
({\it b}) $T$ = $2\times10^6$~K and solar abundances; 
({\it c}) $T$ = $2\times10^6$~K, Ne/H = 10 $\times$ (Ne/H)$_\odot$,
   and solar abundances for the other elements; and
({\it d}) solar abundances and two temperature components,
     $T_1$ = $2\times10^6$~K and $T_2$ = $9\times10^6$~K, 
     with the $T_1$ component five times brighter than the $T_2$
     component.
Note that these model spectra are for comparison only and not the 
results of spectral fits.
}
\end{figure}

\begin{figure}
\epsscale{1.0}
\centerline{\hspace*{2.5cm}\plottwo{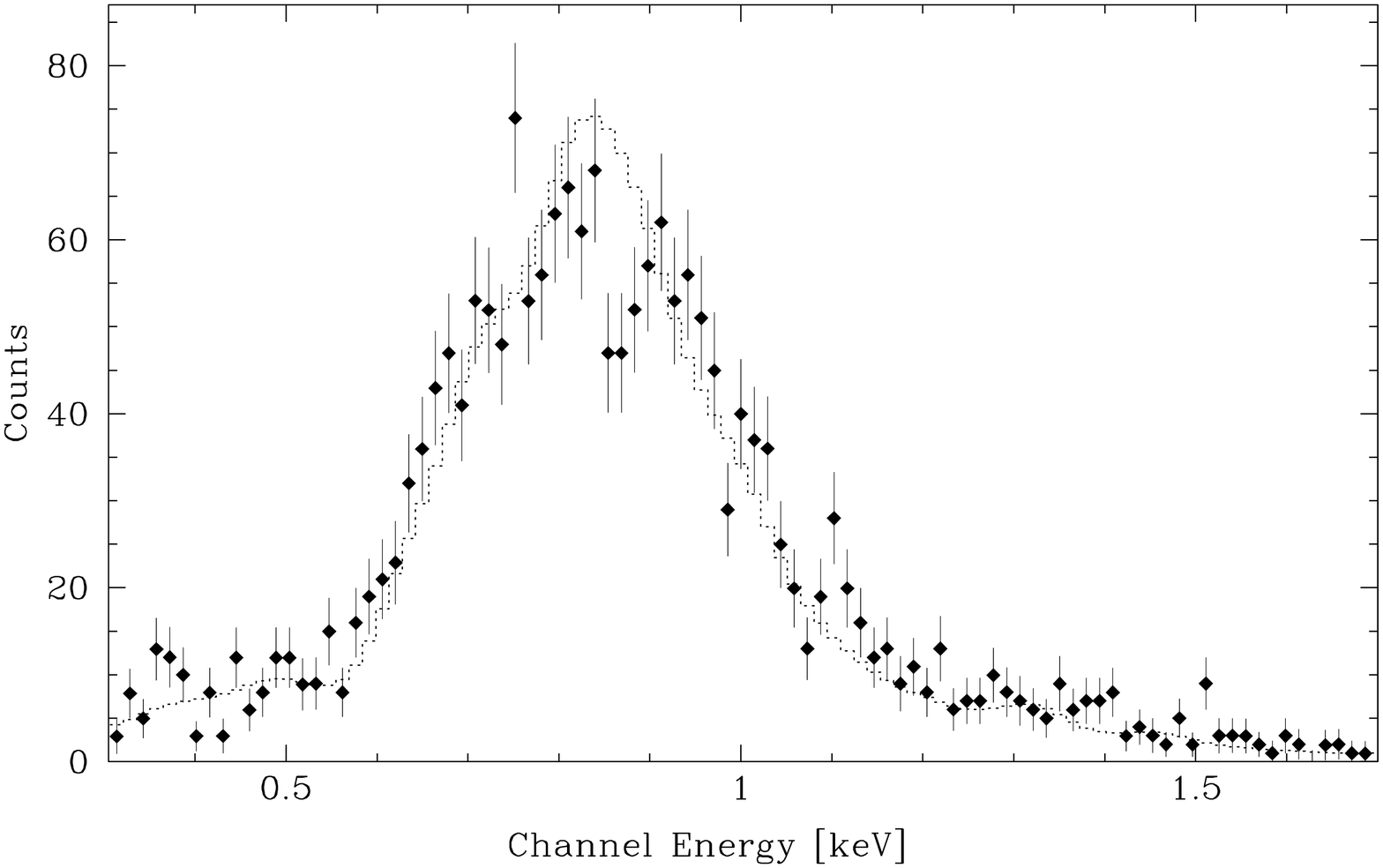}{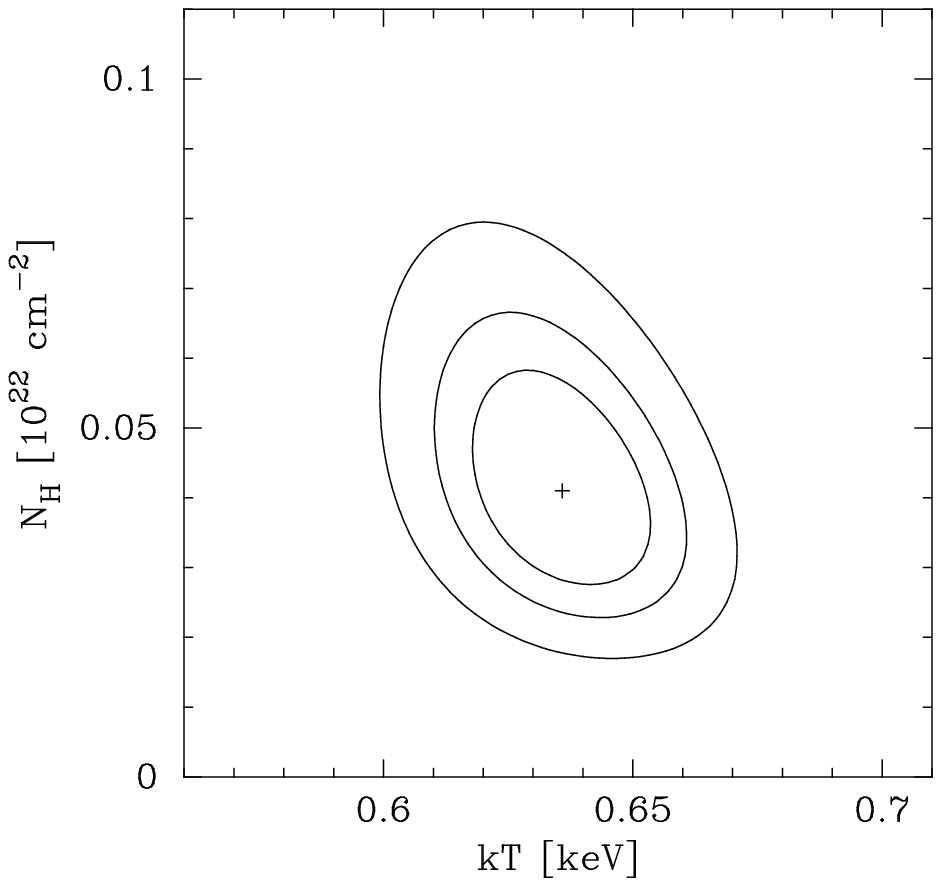}}
\caption{
{\it Chandra} ACIS-S spectrum ({\it left}) and $\chi^2$ grid 
plot of the spectral fit ({\it right}) of the central star 
of NGC\,7293.  
The spectrum is overplotted with the best-fit MEKAL model
that has $kT$ = 0.64 keV and $N_{\rm H}$ = 4$\times$10$^{20}$ 
cm$^{-2}$ (dotted histogram).
The contours in the $\chi^2$ grid plot represent the 68\%, 90\%,
and 99\% confidence levels.  
}
\end{figure}

\begin{figure}
\epsscale{0.6}
\centerline{\plotone{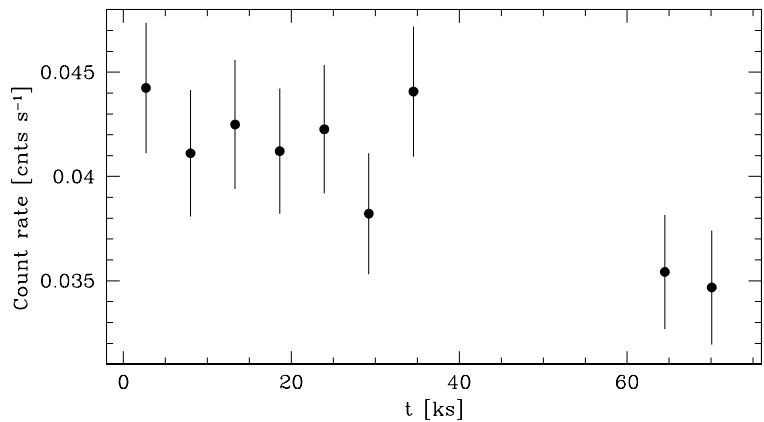}}
\caption{
Light curve of the {\it Chandra} ACIS-S count rate in the 0.3--2.0 keV 
range of the central star of NGC\,7293.  
The observations started at JD 2451499.993.  
}
\end{figure}

\end{document}